\documentclass{article}

\PassOptionsToPackage{numbers, compress}{natbib}
\usepackage[final]{include/neurips_2019}

\usepackage[utf8]{inputenc} 
\usepackage[T1]{fontenc}    
\usepackage{hyperref}       
\usepackage{url}            
\usepackage{booktabs}       
\usepackage{amsfonts}       
\usepackage{nicefrac}       
\usepackage{xfrac}       
\usepackage{microtype}      
\usepackage{wrapfig}

\usepackage{collcell}
\usepackage{xstring}
\usepackage{tabularx}
\usepackage{graphicx}
\usepackage{tabu}
\usepackage{environ}
\usepackage{fp}
\usepackage{upgreek}
\usepackage{amsmath}
\usepackage{amsthm}
\usepackage{enumitem}
\usepackage{bm}
\usepackage{subfiles}
\usepackage{array,multirow}
\usepackage{caption}
\usepackage{subcaption}
\usepackage{makecell}
\usepackage{tikz, color}
\usetikzlibrary{matrix}
\usetikzlibrary{calc}
\usetikzlibrary{positioning}
\usetikzlibrary{graphs}
\usetikzlibrary{shapes.geometric}

\usepackage{xcolor,colortbl}
\definecolor{matlab-blue}{rgb}{0,0.4470,0.7410}
\definecolor{matlab-orange}{rgb}{0.8500,0.3250,0.0980}
\definecolor{matlab-yellow}{rgb}{0.9290,0.6940,0.1250}

\definecolor{matlab-green}{rgb}{0.4660,0.6740,0.1880}

\definecolor{matlab-red}{rgb}{0.6350,0.0780,0.1840}

\title{On the Importance of Opponent Modeling\\in Auction Markets}

\author{
  Mahmoud Mahfouz*$^{1, 2}$, \hspace{0.25em}
  Angelos Filos*$^{1, 3}$, \hspace{0.25em}
  Cyrine Chtourou$^{1}$, \hspace{0.25em}
  Joshua Lockhart$^{1}$ \\
  \AND
  Samuel Assefa$^{1}$, \hspace{0.25em}
  Manuela Veloso$^{1}$, \hspace{0.25em}
  Danilo Mandic$^{2}$, \hspace{0.25em}
  Tucker Balch$^{1}$ \\
  $^1$ JPMorgan Chase \& Co., \hspace{0.25em}
  $^2$ Imperial College London, \hspace{0.25em}
  $^3$ University of Oxford 
 \\
  \texttt{mahmoud.a.mahfouz@jpmchase.com}
  \thanks{Equal contribution}
}

\begin{document}

\maketitle

\begin{abstract}
The dynamics of financial markets are driven by the interactions between participants, as well as the trading mechanisms and regulatory frameworks that govern these interactions. Decision makers would rather not ignore the impact of other participants on these dynamics, and should employ tools and models that take this into account. To this end, we demonstrate the efficacy of applying \emph{opponent-modeling} in a number of simulated market settings. While our simulations are simplified representations of actual market dynamics, they provide an idealized ``playground'' in which our techniques can be demonstrated and tested. We present this work with the aim that our techniques could be refined and, with some effort, scaled up to the full complexity of real-world market scenarios. We hope that the results presented encourage practitioners to adopt opponent-modeling methods and apply them on live systems, in order to enable not only reactive but also proactive decisions to be made.

\end{abstract}
 \section{Introduction} \label{sec:instroduction}

The study of multi-agent systems (MAS) plays an important role in different scientific disciplines, including, but not limited to biology~\citep{hammerstein1994game}, evolutionary psychology~\citep{cosmides1989evolutionary}, economics~\citep{friedman1986game}, and political science~\citep{morrow1994game}. MAS has also been studied in computational finance to develop trading agents~\citep{wellman2011trading} and their associated decision making strategies, taking into account the interactions with the other market participants. In MAS, the agents attempt to maximize a utility function which is typically dependent on the behavior and actions of the other agents in the environment. This suggests the need for agent strategies to be reactive and adaptive to the other agents in the environment.

Game theory~\citep{shoham2008multiagent, wooldridge2009introduction} provides a mathematical formalism for describing the complex behaviors that emerge from the different agent interactions and suggests tools to develop successful strategies. Nonetheless, strong assumptions are traditionally made about having access to reliable models describing the system dynamics, which is invalid even in the simplest of setups. For example, in financial markets designing a strategy that takes into account the impact of its decisions, is as difficult as solving the original problem itself, and heuristic approaches~\citep{gsell2008assessing} are not adaptive hence their quality degrades as the dynamics drift over time. 

In contrast, learning from data~\citep{shoham2008multiagent} is an attractive framework under which assumptions can be relaxed and flexible models can be developed and calibrated to real-world data. In particular, recent advances in machine learning~\citep{goodfellow2014generative, mnih2015human} allow for black-box, task-agnostic, large-scale modeling. However, in all these settings, the presence of multiple learning agents renders the training problem non-stationary and often leads to unstable dynamics or undesirable final results~\citep{balduzzi2018mechanics, foerster2018learning}. 

Opponent modeling~\citep{carmel1995opponent} addresses this problem, attempting to \emph{decouple the sources of non-stationarity} by modeling separately the other learning agents from the stationary components of the environment.

In this paper we explore the applications of opponent modeling techniques in single and continuous double auction markets~\citep{wellman2011trading}, as a proxy for real financial markets. We also discuss potential research topics that address key challenges and opportunities of interest to the financial community.
\section{Background and Related Work} \label{sec:background}
Opponent modeling refers to a set of methods to construct models of the other agents in the same environment with the goal of predicting various properties of interest such as the agents' actions, types, beliefs and goals~\citep{albrecht2018autonomous}. Different approaches are studied in the literature including policy reconstruction~\citep{brown1951iterative, veloso1994planning}, type-based reasoning~\citep{barrett2013teamwork, albrecht2015hba}, classification~\citep{weber2009data}, plan recognition~\citep{carberry2001techniques}, recursive reasoning~\citep{carmel1996incorporating}, graphical models~\citep{howard2005influence} and group modeling~\citep{stone1999task}. 

The financial markets can be thought of as a multi-agent system composed of multiple agents whose different utility functions (\emph{e.g.} maximizing profit or reducing risk exposure) are highly dependent on the behavior and actions of the other market participants. For instance, \citet{yang2012behavior, yang2015gaussian} implicitly perform opponent modeling in a continuous double auction setup by using Inverse Reinforcement Learning (IRL) to characterize the different trading strategies in a limit order book. In their approach, they model the trades placed in the order book as a Markov Decision Process (MDP) and cluster the reward space to differentiate between high frequency, opportunistic and market making strategies.

\section{Opponent Modeling in Auction Markets} \label{sec:opponent-modeling}

We formulate decision-making in a multi-agent system as that of finding an optimal policy in a Markov game (MG)~\citep{littman1994markov}. An MG for $N$ agents is defined by a set of states $\mathcal{S}$, $N$ sets of actions $\mathcal{A}_{1}, \ldots, \mathcal{A}_{N}$ for each agent. The next state is produced according to the state transition function $\mathcal{T} : \mathcal{S} \times \mathcal{A}_1 \times \cdots \times \mathcal{A}_N \mapsto \mathcal{S}$, and at each state, agent $i$ receives reward $r_i : \mathcal{S} \mapsto \mathbb{R}$. Each agent $i$ aims to maximize its own total expected return $R_i = \sum_{t=0}^T \gamma^t r^t_i$ where $\gamma$ is a discount factor and $T$ is the time horizon. Note that the transition function depends on \emph{all} agents' actions, hence each agent has an incentive to model and adapt to the other agents actions too, since they contribute to its own expected return.  We distinguish between two types of policies for solving an MG: 

(1) \emph{implicit-modeling}: the agents implicitly take into account the impact of the other agents via the global state, learning a policy $\pi^{(\text{IM})}_{i}: \mathcal{S} \times \mathcal{A}_i \mapsto [0,1]$; 

(2) \emph{opponent-modeling}: each agent explicitly models the policies of other agents, learning a model $\hat{\pi}_{-i}$ and choose actions according to $\pi^{(\text{OM})}_{i}: \mathcal{S} \times \mathcal{A}_i \times {\hat{\pi}_{-i}} \mapsto [0,1]$, where the subscript $_{-i}$ refers to all agents but $i$. 

Note that the implicit-modeling approach is treating the MG as a single-agent Markov decision process (MDP)~\citep{bellman1954theory}, but there are no guarantees for the existence of a stationary optimal policy since the transition dynamics depend on uncontrollable variables (i.e., all agents but $i$).

An attractive framework for learning policies for sequential decision-making from data and without access to reliable models is model-free reinforcement learning~\citep{sutton2018reinforcement} which is also suitable for solving MGs. For agent $i$, we assume parametric policy $\pi_{\boldsymbol{\theta}_{i}}$, which we train using the simple REINFORCE algorithm~\citep{williams1992simple, sutton2000policy}, directly adjusting the parameters $\boldsymbol{\theta}_{i}$ in order to maximize the objective $J(\boldsymbol{\theta}_{i}) = \mathbb{E}_{s \sim p^{\pi}, a \sim \pi_{\boldsymbol{\theta}_{i}}}[R_{i}]$ by taking steps in the direction of $\nabla_{\boldsymbol{\theta}_{i}} J(\boldsymbol{\theta}_{i})$.

In this section we present two examples of auction games, reformulated as MGs and demonstrate the superiority of opponent modeling over opponent-agnostic approaches. Despite the simplicity and abstraction of these examples, links and relevance are drawn with real financial market applications in Section~\ref{sec:impact}. We build on the analysis discussed in ~\citep{wellman2011trading} for auctions and present two representative examples:

\pagebreak

\subsection{First-Price Sealed Bid Iterated Auctions}

\begin{wrapfigure}{r}{0.33\textwidth}
    \centering
    \includegraphics[width=\linewidth]{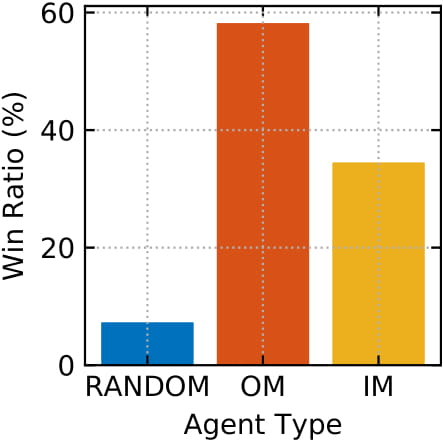}
    \caption{Percentage of auctions won per agent type, where the opponent-modeling (OM) agent dominates both the implicit-modeling (IM) and random policies.}
    \label{fig:win_ratios}
\end{wrapfigure}

In a first-price sealed bid eponymous iterated auction~\citep{wellman2011trading} (also known as a blind auction), all bidding agents simultaneously submit private bids (that is, no bidding agent knows of the other agent's bids) which are revealed at the end of each round. The highest bidder wins, and pays the price they submitted. This process is repeated $T$ times. 

Explicitly, consider $N$ agents, each with a \emph{fixed} and private evaluation $v_i \in [0, 1]$. The $i$-th agent's policy $\pi_{\boldsymbol{\theta}_{i}}$ returns a bid $b_{i} \in [0, 1]$ and the reward received is $v_{i} - b_{i}$ if the agent $i$ won the auction (i.e., submitted the highest bid) or $0$ otherwise. A rational agent does not submit bids larger than its evaluation and tries to maximize the margin $v_{i} - b_{i}$ subject to winning the auction. Note that the game is \emph{stateless}. 

We assume that $N-2$ agents sample from a latent truncated Gaussian distributions with different means $\{\mu_{i}\}_{i=1}^{N-2}$ and variances $\{\sigma^{2}_{i}\}_{i=1}^{N-2}$, maximum value at $\{v_{i}\}_{i=1}^{N-2}$ and minimum value of $0$. 

The remaining two agents are adaptive and trained using the REINFORCE algorithm~\citep{williams1992simple, sutton2000policy}. The first agent (IM) with policy $\pi_{\boldsymbol{\theta}_{\text{IM}}}$, performs implicit modeling. We chose a truncated Gaussian distribution for our model, hence $\boldsymbol{\theta}_{\text{IM}}=\{\mu_{\text{IM}}, \sigma^{2}_\text{IM}\}$ with maximum and minimum values $v_{\text{IM}}$ and $0$, respectively. The second adaptive agent performs opponent modeling and learns a model of the other agents using maximum likelihood estimation for $\hat{\pi}_{-\text{OM}}$ which is used to find the best response $\pi_{\boldsymbol{\theta}_{\text{OM}}}$. 

Figure~\ref{fig:win_ratios} summarizes the results from a simulation with $N=10$, $T=50$ and $v_{\text{IM}} = v_{\text{OM}} > v_{i}$, for all $ i \in \{1, \ldots, N-2\}$. In this experiment, the agent performing opponent modeling wins $58.2\%$ of the auctions demonstrating the value of our approach. The agent performing implicit modeling and the $N-2$ random agents win $34.5\%$ and $7.3\%$, respectively.


\subsection{Continuous Double Auctions}
Previously, we assumed that direct opponent modeling was possible, since the actions of each agent were announced \emph{after} an auction, hence explicit models $\hat{\pi}_{-i}$ could be built on the historic actions. In many cases this information is not available even a posteriori, such as in limit order books of exchanges.

A limit order book (LOB) is used by exchanges to match buyers and sellers of a given security according to price-time priority rules~\citep{bessembinder2001price}. The LOB represents a snapshot of the supply and demand for a given security at a given time. While exchanges maintain historical records of the orders submitted to the limit order book, the identity and the type of the trading participants is not typically revealed. This makes the problem very difficult to approach in a supervised manner using only observations of the order book and the orders placed in the market. 


We create labeled \emph{synthetic} data using an agent-based limit order book simulator (ABIDES)~\citep{byrd2019abides} and formulate our approach to opponent modeling in a limit order book as a supervised learning problem of identifying (classifying) agent \emph{archetypes} explicitly programmed in the simulator, a task which can be cast in the auxiliary tasks framework~\citep{jaderberg2016reinforcement, hernandez2019agent} for semi-supervised reinforcement learning. Using a classification approach for opponent modeling~\citep{weber2009data} allows for predicting properties of the other agents as opposed to predicting the future actions of the agents, which can be useful in a range of market scenarios discussed in our work. 

The trading archetypes considered for generating our synthetic data are: \emph{(1) market makers (MM)}, \emph{(2) liquidity consumers (LC)}, \emph{(3) mean reversion traders (ME)} and \emph{(4) momentum traders (MO)}. For details about the trading agents implemented, please refer to the appendix.

\begin{figure}[!h]
  \centering
  \begin{subfigure}[t]{0.47\linewidth}
    \includegraphics[width=\linewidth]{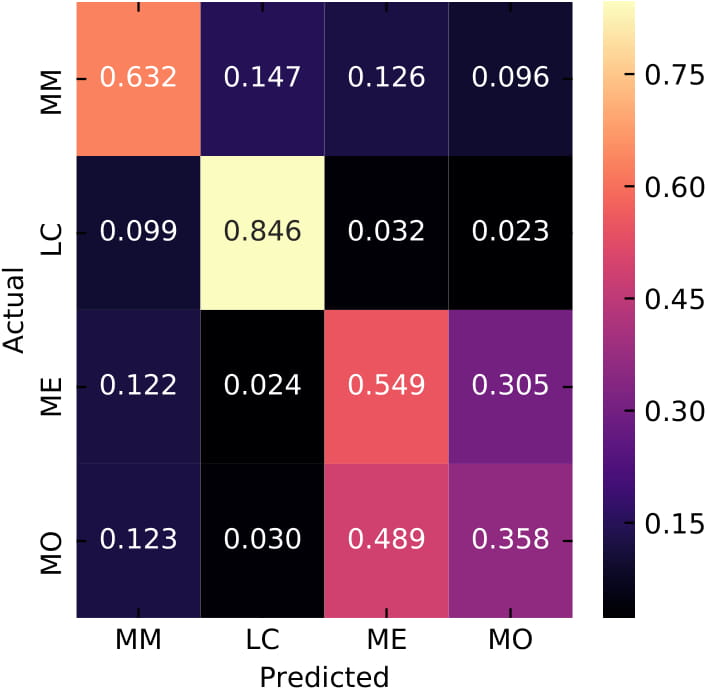}
    \caption{Confusion matrix of a two layer neural network classifier for trading agent types: market maker (MM), liquidity consumer (LC), mean reversion trader (ME) and momentum trader (MO).}
    \label{fig:confusion_matrix}
  \end{subfigure}\hspace{5mm}
    \begin{subfigure}[t]{0.46\linewidth}
    \includegraphics[width=\linewidth]{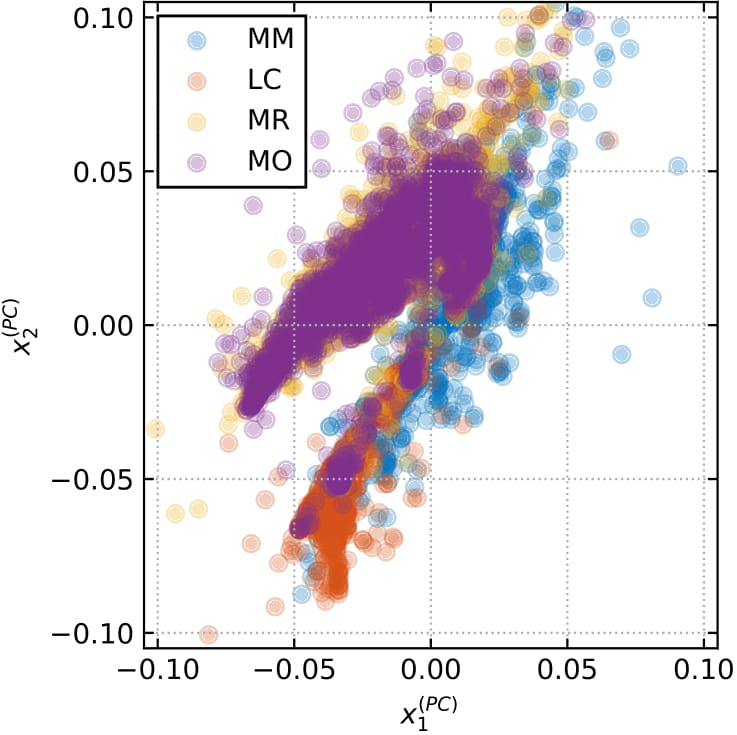}
    \caption{Principal Component Analysis on the feature maps of the penultimate layer of the neural network showing the class separation between the different agent archetypes.}
    \label{fig:pca}
  \end{subfigure}
  \label{fig:double_auction_results}
  \caption{Trading agent archetype classification in a limit order book: model results}
\end{figure}

\vspace{0mm}
\noindent{\textbf{Synthetic Data:}}
The synthetic market is created in an open source agent-based limit order book simulator, ABIDES~\citep{byrd2019abides}. The simulation is run for a trading day between 09:30 AM and 4:00 PM using the specified pool of agents acting against a market replay agent submitting the historical orders for IBM on the 28th of June 2019. The market created consisted of 67 market makers, 500 liquidity consumers, 500 mean reversion traders and 500 momentum agents. An issue arises regarding the different time scales that each agent archetype operates on. For example, some archetypes place substantially more orders than others, leading to the data set we generate having imbalance between the classes. We produce a balanced dataset by simply down-sampling the over-represented classes. The dataset used to train the model consists of the limit order book snapshots representing the state-space, individual agent order details representing the action-space and the agent archetype as the label. The state-space comprises the two best bid and ask prices and sizes in the limit order book prior to the agent's order. We also compute other features including the mid price $M(t)$, the last 5 observed midprices by the agent, the spread (the difference between the best ask price $A(t)$ and best bid price $B(t)$), volume imbalance (ratio of the best ask size to the best bid size) and the order flow imbalance~\citep{cont2014price}, an imbalance metric tracking the net order flow in the bid and ask queues. The action-space contains the order details including the direction of the trade, the price and size of the order, the relative size of the order with respect to the opposite queue and the difference between the order price and the mid price. The synthetic data set created consists of 20,000 state-action pairs (5000 state-action pairs per trading agent archetype).

\noindent{\textbf{Model:}}
The classification model used is a simple feed-forward neural network that attempts to infer the agent archetype given the limit order book state and the agent's order. The model consists of two hidden layers with 128 units each. The activation function used is ReLU with a 0.0004 learning rate and an ADAM optimizer ~\citep{kingma2014adam}. The train-validation-test split used was 53.6 \%, 13.4\% and 33.4\% of the state-action pairs respectively and the model was run for 1000 epochs.

\noindent{\textbf{Results:}}
Using the synthetic data generated and the model discussed, we obtain an 88\% classification accuracy in predicting the different agent archetypes: market maker (MM), liquidity consumer (LC), mean reversion trader (ME) and momentum trader (MO) in the test set. The confusion matrix for the multi-class classification model is shown in Figure~\ref{fig:confusion_matrix}. While our model performs well on classifying the different archetypes, the performance is lower in differentiating between the momentum and mean reversion trading agents. This is expected given the myopic nature of the classifier and the training data used to train the model. In Figure~\ref{fig:pca}, we apply Principal Component Analysis on the feature maps of the penultimate layer of the neural network and project this onto 3 principal components to get an understanding of how the model learns to differentiate between the classes. It is clear that there is a high degree of overlap between the different classes when looking at the first two components which suggests that the problem is not easy to tackle even in the simple setup proposed.
\section{Conclusion and Future Work} \label{sec:impact}

Other than the toy examples demonstrated in Section~\ref{sec:opponent-modeling}, methods of opponent modeling in financial multi-agent systems can be useful in a number of scenarios and from different perspectives, including but not limited to:

\noindent{\textbf{Regulation and Safety:}}
Market manipulation strategies represent an element of price distortion and creation of artificial market conditions, which in turn impedes market efficiency, transparency and fairness. In the LOB context, spoofing agents, for instance, submit orders only to cancel them before their execution and then submit a buy or sell order to benefit from the change that their initial (fake) order caused. The ability to explicitly model the policies of such agents  could help in the early detection of market abuse and avoiding adverse consequences like the flash crash that happened on May 6, 2010, causing a large and quickly reverting dip in the Dow Jones Industrial Average stocks. Moreover, \textit{Level 6} information, consisting of  LOB labeled records, are exclusively available to exchanges and regulators to verify that market participants are complying with rules and processes \citep{paddrik2017effects}. Having access to such a data set is an opportunity to make a more informed and dynamic utility modeling. Quantifying proportions of market makers, liquidity takers and buy side participants would be valuable in detecting signs of recession or unhealthy market conditions.

\noindent{\textbf{Decision Making and Automation:}}
Market participants usually only have access to (at most) \textit{Level 4} LOB data \citep{paddrik2017effects}, which is \textit{anonymized} order flow data with new, modified, and canceled limit orders placed at a given snapshot. Although this poses a challenge to the disambiguation of limit orders placed in the order book because it imposes a myopic approach as opposed to using a consistent timeseries of consecutive orders placed by the same agent, unlabeled data does not fall short of information and opportunities. Opponent modeling can be used for the detection of iceberg orders and other forms of hidden liquidity \citep{hautsch2012dark}. By studying the occurrence of such orders, a \textit{posteriori}, in the LOB and the surrounding market conditions, we can attempt to model the utility behind such strategies. Ref. \citep{hautsch2012dark} studied the correlation between dark orders and market conditions and showed that market conditions reflected by the bid-ask spread, depth and price movements significantly impact the aggressiveness of `dark' liquidity supply. If a trader can detect such dark orders through recognizing a pattern that iceberg order placing agents fail to conceal or through modeling the utility of such agents and the specific cases that make them submit such orders, they can estimate and take into account these orders' effect on the LOB balance and predict a larger than observed spread.  

Additionally, the Request for Quotes (RFQ) process is another setting where we could benefit from opponent modeling. RFQs are used to facilitate trading in illiquid markets \emph{e.g.} fixed income exchange-traded funds (ETFs). In this setting, clients send quote requests to multiple dealers for a specific quantity of a security. Ultimately, the dealer with the most attractive offer wins the auction. In this single, reverse, and static auction setting, one buyer (the auctioneer/the client) is looking to buy a single security from multiple sellers (the dealers) who do not observe the offers submitted by the other dealers, and only receive feedback from the auctioneer once the latter's decision for the winner of the auction is made. It is therefore important for dealers receiving RFQs from different clients to have a good estimate of the expected level at which the other dealers are willing to sell the security to avoid mis-evaluating the offer price. This can in theory be accomplished by having a good estimate of the fair value of the asset in question. However, it is difficult to estimate the fair value of the asset if the asset is illiquid. If a participant is able to infer the utilities, and by extension, the expected offer prices of each of the other dealers in the auction, they can narrow down their uncertainty about the fair value of the asset and avoid mispricing it. In RFQs, apart from modeling competitors, it can be extremely valuable to be able to model clients. In this case data is labeled, as sell side firms (banks) are aware of their RFQ counterparty (client). Performing efficient client tiering would allow to respond to client orders in an efficient manner, which is beneficial for both the client and the bank.

Finally, opponent modeling applied to financial markets, can be a step forward in the automation of decision making. Indeed, explicitly inferring the different market participant strategies would allow to obtain a tailored, well defined utility function that is inspired from high-performing market participants.

We have explored the application of opponent modeling approaches to the problem of building autonomous agents that operate within an auction setting. We have shown how opponent modeling can be applied in the single and double auction settings. In the former case, we showed that reinforcement learning agents that directly model their opponents in a single auction market outperform those that do not. In the double auction setting, we considered the problem of identifying participants in a limit order book from the actions they perform in response to the current state of the order book. We showed that a supervised learning approach can be used to distinguish between different types of market participants in a controlled environment using synthetic data.

The simple experiments explored and the initial results obtained are encouraging evidence that opponent modeling is potentially feasible and useful in settings of interest to the different actors in the financial markets. Future work in this direction could explore more unsupervised or semi-supervised learning approaches especially in the case of the continuous double auction settings. However, the challenges anticipated following such approaches would be around labeling any clusters obtained which would still require domain knowledge. This suggests the need for both supervised and unsupervised learning approaches.

\setcounter{secnumdepth}{0}
\section{Disclaimer}
This paper was prepared for information purposes by the AI Research Group of JPMorgan Chase \& Co and its affiliates ("J.P. Morgan"), and is not a product of the Research Department of J.P. Morgan. J.P. Morgan makes no explicit or implied representation and warranty and accepts no liability, for the completeness, accuracy or reliability of information, or the legal, compliance, financial, tax or accounting effects of matters contained herein. This document is not intended as investment research or investment advice, or a recommendation, offer or solicitation for the purchase or sale of any security, financial instrument, financial product or service, or to be used in any way for evaluating the merits of participating in any transaction.

\bibliography{ms}
\bibliographystyle{plainnat}

\appendix
\newpage
\section{Appendix: Trading Agents} \label{sec:appendix}

\noindent{\textbf{Market Makers:}} These are market participants who aim to provide liquidity in the LOB by placing orders on both sides of the order book. That is, given the market's current mid price $M(t)$, a market maker will place orders at prices $B(t) = M(t) - p_b(t)$ and $A(t)+p_a(t)$ where $p_a(t)$ and $p_b(t)$ are chosen according to some strategy. Market makers operate under the assumption that there is enough trading activity for the instrument that they are operating in, and that the aggregate buying and selling of that instrument will net them a profit proportional to the difference in the prices they offer $A(t) - B(t)$. This difference is referred to as the spread offered by the market maker. In our experiments we utilize two basic market making strategies: The first places two orders (one buy and one sell order) at the top of the book (\emph{i. e. } the highest bid price and lowest ask price in the order book respectively) at each time step. The volumes of the orders are chosen at random following a uniform distribution within fixed bounds $v_\text{min}$ and $v_\text{max}$. The second strategy takes a similar approach but places orders deeper into the book at prices $1,\dots, N$ levels above and below $M(t)$ for $N$ chosen randomly from a uniform distribution within fixed bounds $N_\text{min}$ and $N_{\text{max}}$. In addition, we include an implementation of the Avellaneda-Stoikov strategy \cite{avellaneda2008high}, which dynamically adjusts its bid and ask prices as a function of volatility and the amount of inventory held by the agent.

\noindent{\textbf{Liquidity Consumers:}} In our setup, liquidity consumers represent directional traders placing market orders to buy/sell a given quantity at the best available price. Each agent is configured to place a buy or sell order with the direction and order size chosen at random from a uniform distribution.

\noindent{\textbf{Momentum Traders:}} The momentum agents base their trading decision on observed price trend signals. Our implementation compares the 20 past mid price observations with the 50 past observations and places a buy order of random size, if the former exceeds the latter and a sell order otherwise.

\noindent{\textbf{Mean Reversion Traders:}} These represent agents placing orders in the market following mean reversion signals. In our setup, we have two groups of agents following these type of signals. The first looks at the divergence from an exponentially weighted moving average signal and places buy/sell orders if the price deviates above/below a certain number of standard deviations away from the average. The second group utilizes a similar approach using the Relative Strength Index (RSI) signal \citep{wilder1978new} instead. The RSI is a technical indicator that is used to detect and identify mean reversion. It is defined in terms of the exponential moving average, and the average of up and down price moves over a set period. Although usually the RSI is used on close prices, we will be using a micro-structure variant of RSI where we will look at the time series of mid prices at the tick level. Traditionally, \citep{wilder1978new}, when RSI is greater than the 70 level, the stock is considered to be overbought, and when the indicator is lower than the 30 level, the stock is considered to be oversold. Our agent therefore places a sell order when the stock is overbought, and a buy order when it is oversold and makes no action when the stock is neutral.



\end{document}